\documentclass[manuscript,screen]{acmart}
\usepackage{bbm}
\usepackage{amsmath}
\usepackage{tikz}
\usepackage{verbatim}
\usepackage{caption} 
\usepackage{subcaption}
\usepackage{pgfplots}
\usepackage{adjustbox}
\usepackage[ruled,vlined]{algorithm2e}

\usetikzlibrary{arrows,calc}
\tikzset{
>=stealth',
help lines/.style={dashed, thick},
axis/.style={<->},
important line/.style={thick},
connection/.style={thick, dotted},
}

\AtBeginDocument{%
  \providecommand\BibTeX{{%
    \normalfont B\kern-0.5em{\scshape i\kern-0.25em b}\kern-0.8em\TeX}}}

\copyrightyear{2021}
\acmYear{2021}
\setcopyright{rightsretained}
\acmConference[CHI '21 Extended Abstracts]{CHI Conference on Human Factors in Computing Systems Extended Abstracts}{May 8--13, 2021}{Yokohama, Japan}
\acmBooktitle{CHI Conference on Human Factors in Computing Systems Extended Abstracts (CHI '21 Extended Abstracts), May 8--13, 2021, Yokohama, Japan}\acmDOI{10.1145/3411763.3443428}
\acmISBN{978-1-4503-8095-9/21/05}

\begin{document}
\newcommand{\mysystem}{Bayesian Compass~}
%
\title{Computational Workflows for Designing Input Devices}


\author{Yi-Chi Liao}
\affiliation{%
  \institution{Aalto University}
  \city{Helsinki}
  \country{Finland}
}
\email{yi-chi.liao@aalto.fi}

\renewcommand{\shortauthors}{Yi-Chi Liao}

\begin{abstract}

Input devices, such as buttons and sliders, are the foundation of any interface. The typical user-centered design workflow requires the developers and users to go through many iterations of design, implementation, and analysis. The procedure is inefficient, and human decisions highly bias the results. While computational methods are used to assist various design tasks, there has not been any holistic approach to automate the design of input components. My thesis proposed a series of \emph{Computational Input Design} workflows: I envision a sample-efficient multi-objective optimization algorithm that cleverly selects design instances, which are instantly deployed on physical simulators. A meta-reinforcement learning user model then simulates the user behaviors when using the design instance upon the simulators. The new workflows derive Pareto-optimal designs with high efficiency and automation. I demonstrate designing a push-button via the proposed methods. The resulting designs outperform the known baselines. The Computational Input Design process can be generalized to other devices, such as joystick, touchscreen, mouse, controller, etc. 

\end{abstract}



\maketitle

\section{Introduction}

Input devices, such as buttons, joysticks, sliders, and knobs, are ubiquitous, and they are the fundamental components of any user interface.
Many research has shown that the design of the input devices highly affects user behaviors, performances, and experiences \cite{Kim2014DifferencesIT,ActivationForceTravel}. 
For example, the design of push-buttons results in distinct typing speed and comforts \cite{effectofkeyboard}, and the transfer functions of a mouse lead to different pointing efficiency \cite{gain_pointing}.
However, making a good design of an input device is extremely challenging.
The standard workflow is so-called \emph{User-Centered Design (UCD)}, which usually consists of \textit{design, Implement, and analyze} phases \cite{ucd_paper}.
This process is inefficient and costly.
The designers and developers have to craft several prototypes in order to test them with the users.
User researchers then need to plan and conduct controlled experiments.
The experiment results require interpretations so that the designers can make improvements for the next iteration.
Every step is potentially biased by human decisions. 
For instance, the selection of test sets mostly relies on the designers' experience. 
Because of the high cost, a complete user-centered design process is hardly conducted.
Unfortunately, sometimes designs are fixed based on what feels right to the designers. 
This is even more prone to miss good designs.

\begin{figure*}[t!]
\centering
  \includegraphics[width=0.99\textwidth]{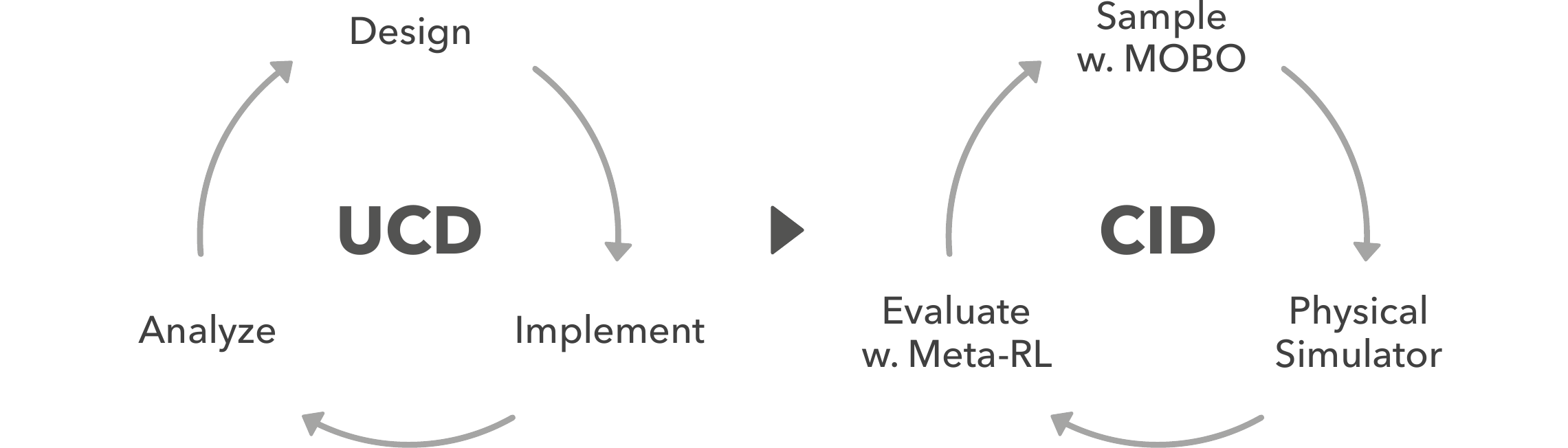}
  \caption{Traditional User-Centered Design (UCD) workflow is inefficient, and the resulting designs are highly biased by the designers' decisions. My thesis proposes \emph{Computational Input Design (CID)}, which contains several components: A multi-objective Bayesian optimization (MOBO) algorithm efficiently searches for Pareto-optimal designs across the design space; A physical simulator instantly renders the design selected by the optimizer; A meta-reinforcement learning (meta-RL) user model adapts to the design instance within a handful of trials and provides performance metrics for evaluating this instance. Jointly, the CID workflows automatically derive Pareto-optimal design instances.} 
    ~\label{fig:teaser}
\end{figure*}

While the same difficulties exist for software-based interfaces, such as menu and layout designs, there are many computational tools for automating the design process \cite{sketch_explorer,menu_optimizer}.
\emph{User models} can be applied for simulating user performances \cite{fitts_law_model,aim_aalto}, that further allow \emph{optimization methods} to derive the optimal designs in a \emph{simulation environment} \cite{7807191, virtual_key_opt}.
However, those computational resources are usually not suitable for input device designs.
First, the construction of user models for physical interaction is generally a challenging task, and those models are typically task-specific.
Secondly, few works have attempted to create input-device simulators that allow realistic rendering of arbitrary designs. 
Thus, only a few papers addressed using optimization methods to automatically search for better input interface design.

I propose \emph{Computational Input Design (CID)} workflows to provide high automation, efficiency, and robust results to the designers (Figure \ref{fig:teaser}). 
Three innovations should be implemented to complete CID: (1) sample-efficient \emph{multi-objective Bayesian optimization (MOBO)} algorithms to efficiently search for Pareto-optimal designs, (2) \emph{physical simulators} for deploying designs quickly without fabricating and assembling an entire physical interface, and (3) \emph{meta-reinforcement learning (meta-RL) user models} to simulate human's ability to quickly adapt to new devices and design instances.

In my thesis, I demonstrate designing a push-button upon the proposed framework.
Buttons are transducers that register a discrete event from physical motion \cite{impact, integration, neuro}, and arguably the most basic input component for any interfaces.
Interestingly, each button design is unique in its haptic response characteristics. 
Gamers, programmers, and hobbyist groups have a keen interest in tactility, which is associated with sensory experience and performances. 
The design of push-button is a costly process where the developers have to craft physical prototypes for further conducting studies. 
The CID process can greatly reduce the cost. 
In my previous works \cite{buttondesign, liao, pressem}, I introduced a novel FDVV models to parameterize button design. 
A sophisticated physical simulator is introduced to render arbitrary tactility according to the given design parameters.
A MOBO algorithm is applied to efficiently explore the design space and search the Pareto-optimal designs.
The user model based on meta-RL algorithms is trained on various example buttons that allow fast adaptation to new design instances.

\subsection{Contribution}
I have introduced a novel FDVV button model allowing realistic button simulation and interactive design \cite{buttondesign, liao, pressem}.
As shown in Figure \ref{fig:pipeline}, an end-to-end simulation pipeline covers capturing the button tactility, modeling, controlling, and rendering.
The guidelines of this work can go beyond buttons to general haptic-rendering devices.
The other paper, MOBO for user interface design, is under submission (an overview of the algorithm is illustrated in Figure \ref{fig:bayes_teaser}). 
In this work, I introduced a novel workflow allowing designers to efficiently identify Pareto-optimal design candidates using Bayesian optimization. 
I demonstrated and assessed the MOBO workflow across three representative interface design problems and the resulting designs outperform all the baselines.
The final component of my dissertation is a meta-RL-based model able to quickly adapt to various input components, including the unseen design instances.  
This paper will close the loop of \emph{Computational Input Design} process.
The contributions of my thesis are as follow:
\begin{itemize}
    \item Proposing \emph{Computational Input Design} workflows and identifying the major components to realize the concept, which further allows a highly efficient and robust input design process;
    \item Three novel implementations: the multi-objective Bayesian optimizer, the physical simulator, and the meta-RL algorithm, can be useful for other general design tasks;
    
    \item Demonstrating designing a button via the CID workflows.
    
\end{itemize}

\subsection{Background}
I am a third-year Ph.D. student in the School of Electrical Engineering, Aalto University, working on Computational Interaction, advised by Professor Antti Oulasvirta.
My research focus is building input devices and touch interfaces utilizing computational methods, such as modeling, Bayesian optimization, reinforcement learning, and meta-learning. 
My ultimate goal is to bring HCI one step closer toward a principled design process and optimized physical interactions.

\section{Key Related Works}

Computational interaction is one branch of the modern HCI field where researchers strive for constructing user models and applying optimization methods to automatically find optimal designs \cite{oulasvirta2018computational}.
Here, we review the development of computation interaction and its limitation on input devices. 
Then, we discuss the methods applied in my thesis.

\subsection{Computational Methods on Input Device Design}

In contrast to the traditional User-Centered Design research process\cite{ucd_paper}, computational interaction aims for improvements in modeling, optimization, and inference. 
Better user models allow more precise prediction of user behavior and performances \cite{fitts_law_model}.
Promising optimization methods enable faster search for the optimal designs \cite{oulasvirta2020combinatorial}. 
Those computational tools have been applied on various kinds of interfaces, such as sketch, menu, and layout design, visualization of scatter plots, etc. \cite{sketch_explorer,menu_optimizer,7864468}.
However, only a few computational methods are applied for assisting input component design.
Some notable exceptions include tuning the gain function \cite{autogain} and investigating the optimal design of mouse sensor position \cite{mouse}. 
These works only applied optimization algorithms for selecting parameters, but no user models nor configurable physical components were involved. 
My thesis is, for the first time, proposing a holistic Computational Input Design framework in which models directly interact with the optimizer and the physical simulator.

\subsection{Bayesian Optimization} 

A part of our paper is built upon Bayesian optimization, a machine learning method that aims to efficiently explore a black-box function and further identify an optimal point. 
The approach is suitable for applications in which the functions are expensive to evaluate in terms of the required time or effort \cite{shahriari_taking_2016}. 
Bayesian optimization has been applied in games to assign parameter values to maximize the user engagement \cite{khajah_designing_2016}.
Conducting crowd-sourcing Bayesian optimization studies is another approach to minimize the required time period \cite{khajah_designing_2016,dudley_crowdsourcing_2019-1}.
\citet{brochu_bayesian_2010-1} demonstrated a technique for allowing designers to quickly determine appropriate values for animation rendering.
Koyama et al. \citep{koyama_sequential_2017-1, koyama_sequential_2020} attempted to assist users on editing of photographs to achieve a promising visual appearance.
Bayesian optimization has also been used for customizing interface settings for individuals \cite{nielsen_perception-based_2015, snoek_bayesian_2013}.
However, the application of Bayesian optimization to HCI design problems has largely been limited to the optimization for tasks with only a single objective.
In practice, most interface design problems are characterized by a complex interplay between multiple, often competing, objectives, e.g., speed versus accuracy.
In my thesis, I leverage the MOBO algorithms introduced by \citet{shah_pareto_2016} and demonstrate its applicability to input device design process.

\subsection{Physical Simulator and Haptic Rendering}

A part of my thesis is built upon physical simulators that aligned with haptics research pursuing the creation of rich and realistic sensations~\cite{worldoftouch}. 
Research has looked at advanced factors affecting haptic perception, such as friction and texture \cite{softness-display}. 
Force simulators have been introduced, for example, the Phantom device is a 6-DOF pen-type force-rendering device capable of emulating the softness of deformable objects \cite{thephantom94}. 
However, a low operating rate (60 Hz), excessive degrees of freedom, and a lack of vibrotactile stimulus limit its use to simulate input devices accurately.
Doerrer and Werthschuetzky enabled users to edit the force-displacement profile of a push-button in software \cite{Doerrer02}. 
Nonetheless, the force-displacement model is known to be incomplete, and there are no experimental results to validate this approach.
Hence, there is a need to explore more effective methods for constructing physical simulator and haptic rendering.

\subsection{Meta-Reinforcement Learning Algorithms} 
Lastly, the user model in my thesis is largely based on meta-reinforcement learning (meta-RL) algorithms.
Reinforcement learning (RL) is an area of machine learning concerned with how software agents take actions in an environment in order to maximize the cumulative reward function \citet{sutton2018reinforcement}.
Recently, various approaches have been introduced for reinforcement learning with neural network function approximators. 
The well-known algorithms are deep Q-learning \cite{mnih2015human}, trust region policy gradient methods \cite{schulman2017trust}, proximal policy optimization \cite{schulman2017proximal}, and soft actor-critic \cite{haarnoja2018soft}. 
On the other hand, Meta-Learning concerns the question of learning to learn. 
The goal of meta-learning is to train a model that can quickly adapt to a new task using only a few data points and small training iterations \cite{duan2016rl,finn2017modelagnostic}. 
The model is previously trained during a meta-learning phase on a set of similar tasks, such that the trained model can quickly adapt to new unseen tasks.
There are several variation of meta-RL implementations, such as LSTM meta-learner \cite{ravi2016optimization} and gradient-based model-agnostic meta-learners \cite{finn2017modelagnostic, nichol2018firstorder}.
A recent development is based on gradient-based meta-learner combined policy-gradient methods shows a good fit for continuous state, action environments \cite{rothfuss2018promp}.
I plan to follow this approach to build a meta-RL model to simulate continuous user's hand motions.
\section{Research Progress}
\label{current}

My current publications have contributed to the vision towards state-of-the-art MOBO for interaction design and realistic button simulation.

\subsection{Multi-Objective Bayesian Optimization for Interface Design}

\begin{figure*}[t!]
\centering
  \includegraphics[width=0.99\textwidth]{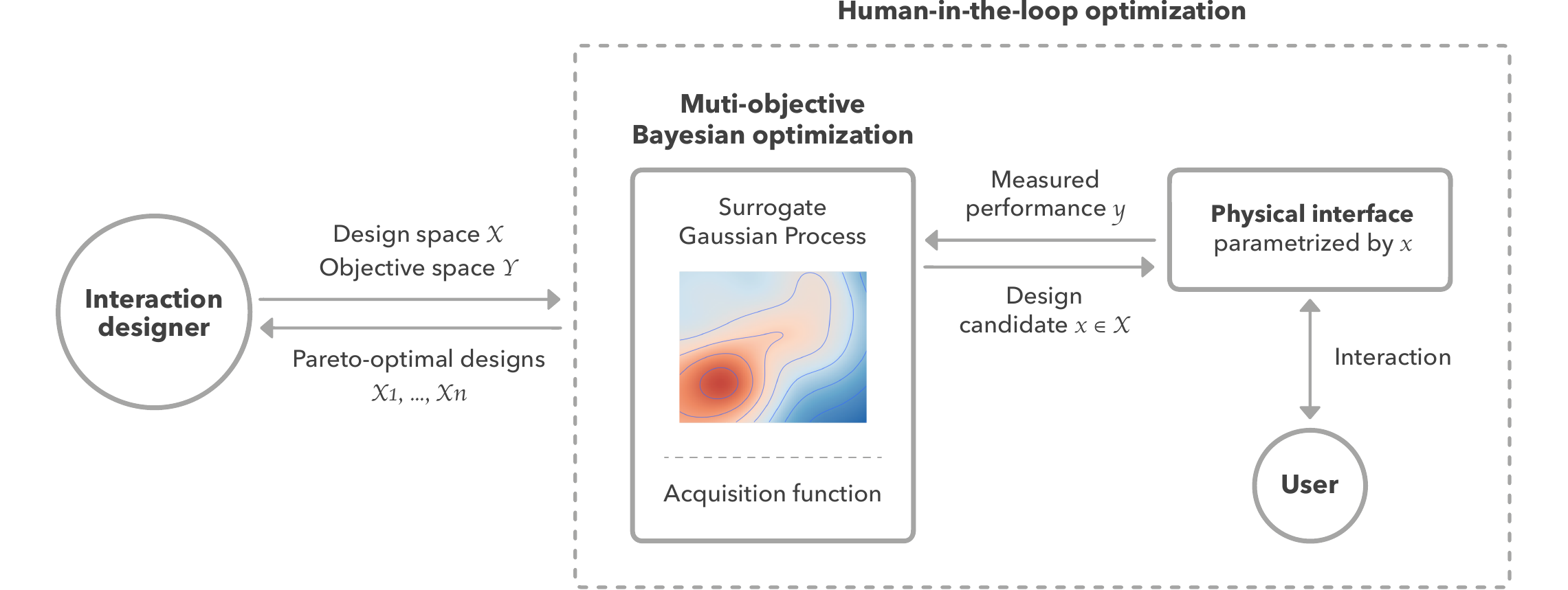}
  \caption{Multi-objective Bayesian optimization (MOBO) is a novel workflow for efficiently identifying a set of informative designs. The designer defines design parameter space ($X$) and objective function space ($Y$). 
  The optimizer will generate design candidates $x\in X$ to search for the Pareto-optimal points. The interface will render according to each $x$. The user's interaction will then be translated into objective values $y$. The optimizer updates its proxy model, a Gaussian Process model defined by the observed $\{x,y\}$ sets. Then, the proxy model generates a new design candidate, and sends it to the interface.} 
    ~\label{fig:bayes_teaser}
\end{figure*}

I introduced Bayesian optimization as an efficient solution to multi-objective design challenges related to interaction techniques (Figure \ref{fig:bayes_teaser}).
Specifically, I described and demonstrated an interactive workflow for identifying \emph{Pareto-optimal design candidates} for HCI design tasks using Bayesian optimization.
I extended \emph{Bayesian optimization} \cite{shahriari2015taking}, a powerful approach for efficient identification of optimal designs in studies involving human data.
The value of the approach lies in its efficient exploration of the design space, with the technique particularly well suited to black-box optimization with noisy observations and its evaluation is expensive.
An intuitive way to think about it is that it can help the designers to \emph{avoid} testing design instances that would be uninformative, for instance, parameters associated with poor usability or some measurable objectives.
I demonstrated using the MOBO workflow for tackling representative input device design problems.
In all case studies, the results indicated that MOBO design approach enables the efficient identification of promising design candidates.

\subsection{Button Simulation and Design}

\begin{figure*}[t!]
  \includegraphics[width=0.99\textwidth]{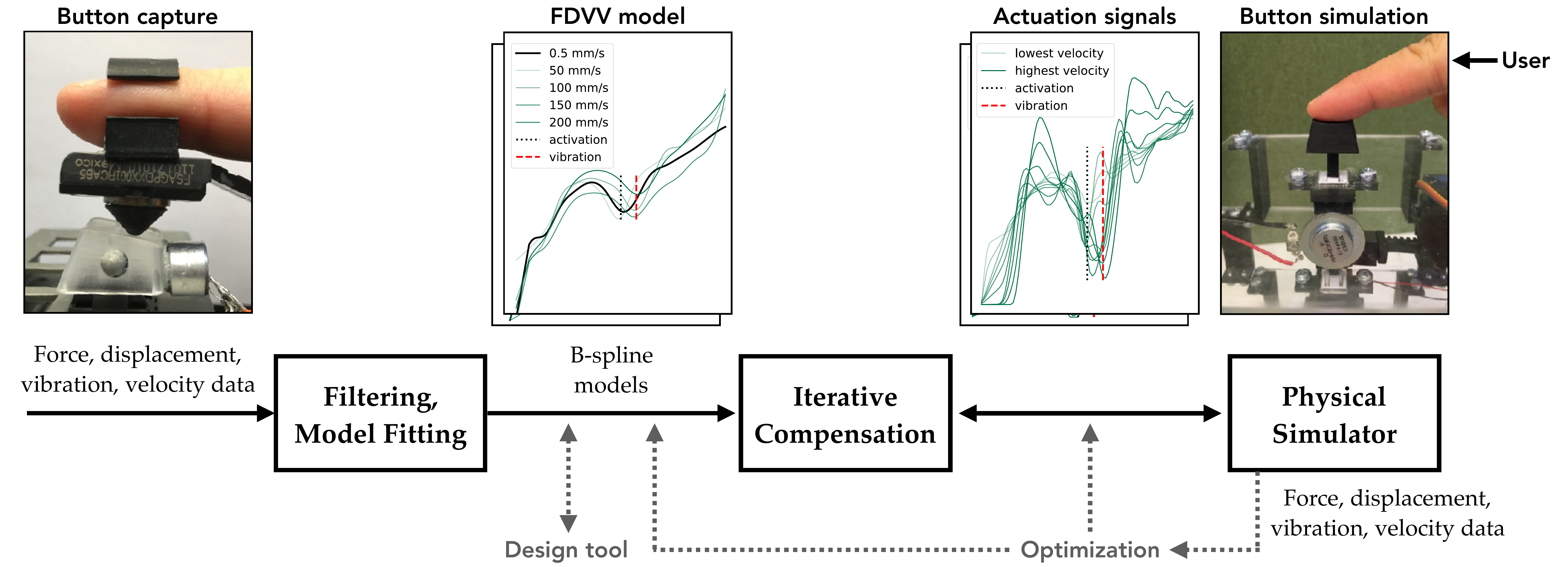}
  \caption{An overview of my previous works on button simulation and design \cite{buttondesign, liao,pressem}. In this series of works, I introduced an end-to-end approach for button simulation and design. To capture an FDVV model of a button, sensors are placed on the finger, and the button is pressed multiple times. The resulting force, displacement, vibration, and velocity data are filtered and fitted with the lower-parametric B-splines models based on the Bayesian Information Criteria (BIC) values. A designer can edit the model produced. Before rendering the FDVV model on our simulator, an iterative compensation process computes how to cancel the simulator's transfer function. The resulting actuation signals drive the simulator.
  }
  \label{fig:pipeline}
\end{figure*}

My previous publication investigated the simulation and interactive design of push-buttons \cite{buttondesign, liao,pressem}. 
Accurate simulation of the button-pressing experience is challenging. 
While pre-existing simulators can render tactile and linear buttons, I found that the typical force--displacement (FD) model cannot accurately render button tactility.
Also, no methods have been offered to help designers and engineers exploit such simulations.
To address these challenges, I proposed an extended model and an end-to-end simulation pipeline around it (Figure \ref{fig:pipeline}).
This approach allows simulating more button types than previously, including tactile-type buttons and buttons with different click reactions and various travel ranges.
Furthermore, it permits the analysis and editing of buttons. 
Our work centers on the \emph{Force--Displacement--Vibration--Velocity} (FDVV) model and an end-to-end simulation pipeline.
The model adds vibration response and velocity-dependence on top of the FD model. 
In our implementation, vibration is sampled through a microphone during a button press, and multiple FD curves are sampled at several speeds.
I solved several engineering challenges connected with ambitions to capture and simulate buttons via FDVV models. 
For rendering it, I presented a novel simulator construction for FDVV models. 
This simulator is capable of detecting displacement to $\mu$m precision at a high sampling rate (1~kHz) and can produce a wide range of force (up to 4.4~N) and vibration (50~Hz~--~20~kHz) feedback. 

\section{Future Directions}

As part of future work, I am working on a general user model that can learn policies for interacting with physical devices and quickly adapt to new design instances.
Such a user model can be used for simulating user behavior and thus reduce the efforts of conducting rounds of user studies.
For modeling interactions with input devices, such as pressing a button, reinforcement learning is so far the most general and prominent framework \cite{sutton2018reinforcement}. 
Meta-RL further provides a means that allows the model to quickly identify an optimal policy for a new task by leveraging the previous learning experiences.
The meta-RL models bring us one step closer to the human-like behaviors when encountering a new input device.
While various meta-RL algorithms have been introduced, a particular algorithm may be suited to the needs of modeling continuous human hand motions, which is based on gradient-based meta-learning and policy-gradient methods \cite{rothfuss2018promp}. 
I intend to follow this approach to build a general-purpose user model.
The goals of this paper are: (1) constructing a general-purpose model for the physical interaction of wide-ranging input devices, and
(2) by merging the meta-RL model into other proposed workflows, we can learn the efficacy of a fully automated \emph{computational input design} process.
The first design task is deriving a push-button via the CID workflows: 
The physical button simulator renders button design based on the MOBO's suggestion (as mentioned in section \ref{current}), and a meta-RL model applied on a robot will act as various users to interact with the button simulator. 
Jointly, a series of Pareto-optimal button designs will be derived automatically. 
Other input devices, such as joystick and mouse, can potentially be designed by the proposed CID process.
\section{Acknowledgement}

I would like to thank my colleagues and collaborators from Aalto University, DGIST, KAIST, University of Cambridge, National Yang Ming Chiao Tung University, and my advisor Antti Oulasvirta for their guidance and support.


\bibliographystyle{ACM-Reference-Format}
\bibliography{acmart}


\end{document}